\documentclass[12pt,dvips]{article}
\newcommand{\dg}[1]{\mbox{$\rlap /{\! #1}$}}
\newcommand{\eps}{  \!\!\mbox{ \large$\epsilon$ }\!\!  }

\newcommand{\abs}[1]{  \mbox{ $\left|{#1}\right|$ }  }
\newcommand{\xp}[1]{ \mbox{ $\mbox{\large e}^{#1}$ } }
\newcommand{\dt}{ \mbox{ $\!\!\cdot\!\!$ } }
\newcommand{\dfrac}[2]{  \mbox{ ${\displaystyle\frac{#1}{#2}}$ }  }
\usepackage{amstex}
\usepackage{graphicx}
\def\BM#1{{\bf#1}}

\newcommand{\cll} {\mbox{$C_{\lambda\lambda'} $}  }
\newcommand{\ub}  {\mbox{$\bar{u}(p,\lambda)  $}  }

\newcommand{\upp}  {\mbox{$u(p',\lambda')      $}  }

\newcommand{\vp}   {\mbox{$\vec{p}             $}  }
\newcommand{\vpp}  {\mbox{$\vec{p}{}\ '            $}  }
\newcommand{\ps}   {\mbox{$\rlap /\;\!\!p      $}}

\newcommand{\betta}   {\BM{\eta}}

\begin{document}

\begin{flushright}
SMU-HEP-94/28 \\ UCRHEP-T152 \\[12pt] \today
\end{flushright}

\vfill
\vspace{24pt}
\begin{center}

{\bf A Covariant Method for Calculating Helicity Amplitudes}

\vspace{12pt}

Roberto Vega~\footnote{vega@@mail.physics.smu.edu}\\
{\sl Department of Physics}\\
{\sl Southern Methodist University}\\
{\sl Dallas, TX 75275}

\vspace{10pt}

Jose Wudka~\footnote{jose.wudka@@ucr.edu}\\
{\sl Department of Physics}\\
{\sl University of California at Riverside}\\
{\sl Riverside, CA 92521-0413}

\vspace{12pt}

\begin{abstract}
We present an alternative  approach for calculating
helicity amplitudes for processes involving both massless and massive
fermions.    With this method one can easily obtain covariant
expressions  for the helicity amplitudes.  The final expressions involve
only four vector products and are independent of the basis for gamma
matrices or specific form of the spinors.   We use the method to obtain
the helicity amplitudes for several processes  involving top quark
production.
\end{abstract}
\end{center}

\vfill
\pagebreak
\setcounter{page}{1}

\section{Introduction}

The conventional approach used to determine decay rates and cross
sections  is to square the Feynman amplitude, average over initial, and
sum  over final spin states.  This approach is very inconvenient to
implement if the number  of Feynman diagrams and the number of final
state particles is large.  This is true even with the use of all the
modern tools for symbolic  computation.   The conventional approach also
has the drawback that in  summing over spin states one loses information
on spin correlations.  Such correlations may be important for exploring
new physics at future  colliders.

An alternative approach, which has gained popularity in the past
decade, is to compute the helicity amplitudes numerically.  By doing
this, spin correlations are easily explored and the process of
squaring the amplitude is trivial.  The idea of calculating helicity
amplitudes is probably as old as the conventional approach and might
even pre-date it\cite{jacob/wick-59,bj/chen-66}.  More recently,
motivated by the need to consider more complicated processes, a number
of novel methods\cite{calcul-81,gun/kun-85}
\nocite{calcul-82,calcul-82b,calcul-82c,calcul-84a,calcul-84b} for
obtaining helicity amplitudes have been developed. Most of these,
however, are viable only in the limit where fermion masses can be
ignored.  In this limit helicity and chirality are equivalent, leading
to enormous simplifications; this is particularly evident in processes
involving the $W$-gauge bosons. Most of the recent so-called helicity
methods\cite{calcul-81,gun/kun-85, farrar/neri-83} take advantage of these
simplifications\cite{haber-94}.

Although, there have been several attempts to generalize the helicity
methods for massless fermions to the massive case, the resulting
procedure is rather cumbersome and requires the introduction of
arbitrary reference momenta\cite{kleiss/stirl-85}. It is the purpose of
this paper to present a simple and systematic method for calculating
helicity amplitudes which works for processes involving both massless
and massive fermions.   The approach that we present leads to a
covariant expression for the helicity amplitude in terms of a minimum
number of four vector products and is independent of the basis for Dirac
matrices or specific forms of the spinors.   Thus, we believe that our
approach leads to an analytic expression which is relatively simple and
easy to compare with.

For illustration we apply our method to compute the helicity amplitudes
for various processes involving the top quark.  We consider the processes:
\begin{gather*}
e^+e^- \rightarrow t\bar{t} \\
\gamma\gamma  \rightarrow t\bar{t}
\end{gather*}

\section{The Method}

When evaluating a Feynman amplitude involving a pair fermions and several
gauge bosons, in the initial or final state, the amplitude is expressed
as a sum of terms which have the form,
\begin{equation}
\bar{u}(p,\lambda )\,\dg{B}_0\dg{B}_1\dg{B}_2\cdots\dg{B}_{2n+1}\,
u(p',\lambda^\prime ), \label{one}
\end{equation}
where $(p,\lambda)$ and $(p',\lambda^\prime)$
are the momenta and helicity of the external fermions.
The particular labeling of the indices is chosen for convenience
in the manipulations that follow.  Note that, as written, (1) has
an even number of $\gamma$-matrices sandwiched between spinors.
The case where the number of $\gamma$-matrices is odd is treated
by setting $\dg{B}_0=I$.

In the usual approach one squares these terms and sums over the fermion
spins to obtain a sum over terms of the form,
\begin{equation}
tr[(\ps+m)\,\dg{B}_{2n+1}\dg{B}_{2n}\cdots
\dg{B}_1\dg{B}_0\,(\rlap/{p'}+m')\,
\dg{B}_0\dg{B}_1\dg{B}_2\cdots\dg{B}_{2n+1}].
\end{equation}
The traces can then be evaluated and the result will be
a function of Minkowski products of the external particle four
momenta.  The expression is then easily integrated
over phase space by use of Monte Carlo methods.
In this approach one does not need explicit values of components
of free spinors (which depend on the particular representation
used for the $\gamma$-matrices).  This method is also
convenient because, in addition to getting rid
of the $\gamma$-matrices and expressing the amplitude in terms of
simple ``dot'' products, the final result is independent of the
unobservable fermion spins.

In complicated cases where the amplitude involves the sum over a large
number of Feynman diagrams, and each diagram involves many fermion-gauge
boson vertices, the number of traces and the length of the traces may
render the standard approach impractical. The standard
approach is also inconvenient in cases where interference effects are of
interest and one may want to explicitly display the relative phases
between the different Feynman diagrams. In these cases it is more
convenient to obtain an expression for the amplitude corresponding to
each graph which is simple to
evaluate numerically.  The square of the amplitude is then obtained by
just squaring a single complex number.

For processes that only involve massless or nearly massless fermions
there have been many papers where the so called helicity methods for
evaluating tree level amplitudes are developed\cite{calcul-81,gun/kun-85,
farrar/neri-83}.  In
contrast to the standard approach, the amplitude in the Helicity Method
is expressed in terms of spinor products of the form
\[ s(p,p')=\bar{u}(p,+)u(p',-). \]  In this
definition the $p$ and $p'$ are light-like four-vectors. The $\pm$ refer
to the helicity of the fermion.  Note that $\abs{s(p,p')}^2=p\dt p'$, so
that up to a phase $s(p,p')$ contains essentially the same information as
$p\dt p'$. These methods prove relatively simple to implement and lead
to compact expressions valid whenever the fermion masses can be ignored.
Generalizations of this approach to the massive fermion
case\cite{kleiss/stirl-85,ballestrero/maina-94}
exist, but the results are cumbersome: the method
requires the introduction of extra light-like momenta and
the expression of the spinor for a massive particle as a sum over spinors
for massless particles. Consequently, as the number of external particles
increases, the number of terms for each diagram grows and results
quickly become unmanageable and more prone to human error.
In another approach\cite{Barger-p-s-91} some authors found it more
convenient to obtain the amplitudes by performing all the
$\gamma$-matrix products numerically.  Although this approach appears to be
efficient, the result is not expressible in a simple compact analytic
form which one can study. We finally note that all the approaches
quoted above require writing down particular expressions for the
spinors.

We will see below that with the approach presented here, it is possible to
actually write the amplitude in terms of ``dot'' products without the
use of particular expressions for the spinors.  The procedure is
relatively simple and  leads to compact analytic expressions
which are easy to compare with and put into code to evaluate
numerically or symbolically.

The basic idea
is to observe that (1) can be re-written as the
trace,
\begin{equation}
  tr[ u(p',\lambda^\prime ) \otimes
  \bar{u}(p,\lambda )\dg{B}_0\dg{B}_1\dg{B}_2\cdots
  \dg{B}_{2n+1} ].
\end{equation}
Although this trivial observation has been used by
several authors in the past\cite{caffo-82,passarino-83}.  there was
no attempt was made to make use of the Michael-Bouchiat identity
(as we do below) to obtain an expression for
$u(p',\lambda^\prime)\bar{u}(p,\lambda)$.
The object, $u(p',\lambda^\prime ) \otimes \bar{u}(p,\lambda )$,
is a four by four matrix in Dirac space and can be expressed
in terms of an orthogonal basis of the four-dimensional Dirac
space~\cite{eran}.
We do not follow this approach here because we found it simpler
to make use of a spinor identity due to Michel and
Bouchiat~\cite{bouchiat/michel-58},
\begin{equation}
  u(p,\lambda^\prime )\bar{u}(p,\lambda ) =
  \frac{1}{2}(\rlap/{p}+m)
  (\delta_{\lambda\lambda^\prime }
  +\gamma_5\rlap/ \eta^i\sigma^i_{\lambda\lambda^\prime}).
\label{emb}
\end{equation}
The $\sigma$'s are the three Pauli matrices and there is an
implied sum over $i={1,2,3}$.  Note that in the case where $\lambda{}'=\lambda$
expression~(\ref{emb}) reduces to the usual projection operator for a state
of momentum $p$ and helicity $\lambda$.  The $\eta$'s
are defined such that,
\begin{eqnarray}
   \eta_i\dt \eta_j&=&-\delta_{ij}   \nonumber \\
   \eta_i\dt p &=&0.
\end{eqnarray}
They are the spin vectors corresponding to the four momentum $p$.  For,
\[\vec{p}=\abs{\vec{p}}(\sin\theta \cos\phi,\sin\theta\sin\phi,\cos\theta),\]
a standard representation of the $\eta$'s is given by,
\begin{eqnarray}
 \eta_1^\mu&=&(0;\cos\theta\cos\phi,\cos\theta\sin\phi,-\sin\theta)
    \nonumber  \\
 \eta_2^\mu&=&(0;-\sin\phi,\cos\phi,0)     \nonumber   \\
 \eta_3^\mu&=&( \frac{ \abs{\vec{p}} }{m} ;
 \frac{E}{m}\frac{\vec{p}}
 {\abs{\vec{p}}}),
\label{polvec}
\end{eqnarray}
where $\eta^\mu_k=(0;\delta^i_k)$ for the case when $\abs{\vec{p}}=0$.

For completeness we have included a proof of identity (\ref{emb})
in the appendix.
The identity is only valid if both spinors correspond to the same
momenta.  To make use of it we must generalize it to obtain an
expression for $u(p',\lambda^\prime)\bar{u}(p,\lambda)$.   If $m'=m$
this could be done by applying a boost operator to both sides of
(\ref{emb}). It
is, however, simpler and more general to observe that any spinor
$u(p',\lambda')$ can be expressed in the form
\begin{equation}
    u(p',\lambda') = {\cal{N}}\sum_\lambda C_{\lambda'\lambda}
                     (\ps'+m') u(p,\lambda),
    \label{upeqn}
\end{equation}
where  $p'{}^2=m'{}^2$, and $\cal{N}$ is a normalization factor.
This expression
is valid whether or not $m'=m$.  Note that the helicity index
$\lambda$ is defined with respect to the direction of \vp, while
$\lambda'$ is defined with respect to \vpp; i.e. the frames in which
$\lambda$ and $\lambda'$ are defined are rotated with respect to each other.
Multiplying (7) on the right by $\bar{u}(p,\lambda)$ and using identity (4)
we obtain,
\begin{equation}
   C_{\lambda'\lambda}= {\cal{N}}\, \bar{u}(p,\lambda)u(p',\lambda'),
   \label{clleqn}
\end{equation}
with,
\[ {\cal{N}} = \dfrac{1}{\sqrt{2(p\dt p'+mm')}}. \]
It is easy to verify that the matrix $C_{\lambda'\lambda}$ is unitary in
helicity space as it should be.  Using (\ref{upeqn}) in (4) we obtain
the desired generalization,
\begin{equation}
u(p ',\lambda^\prime ) \otimes \bar{u}(p,\lambda ) =
\frac{{\cal N}}{2}
\sum_{\lambda''}\,C_{\lambda'\lambda''}\,
(\ps'+m')(\ps +m)
(\delta_{\lambda\lambda'' }
+\gamma_5\rlap/ \eta^i\sigma^i_{\lambda\lambda''}).
\label{meqn}
\end{equation}
Since $\cll$ is unitary it drops out of our expressions upon summing
the square of the helicity amplitudes over all helicities.
Thus, if in the end one is going to sum over all helicities
one can simplify (\ref{meqn}) by setting
$C_{\lambda'\lambda''}=\delta_{\lambda'\lambda''}$.  The resulting
amplitudes will not correspond to the correct helicity amplitudes
but the process of squaring them and summing over all helicity indices
will indeed generate the correct result for the sum of the squares.

If spin correlations are of interest then one needs to make use of
the specific expressions for the $\cll$.  The expressions follow from
(\ref{clleqn})
\begin{eqnarray}
    C_{++} & = & \left[
    \dfrac{EE'-\abs{\vp}\abs{\vpp}+mm'}{p\dt p'+mm'} \right]^{1/2}
    [\cos{ \frac{\theta}{2} }\cos{ \frac{\theta'}{2} }
  + \sin{ \frac{\theta}{2} }\sin{ \frac{\theta'}{2} }\xp{i(\phi'
    -\phi)}]  \nonumber \\[12pt]
  C_{-+} & = & \left[
   \dfrac{EE'+ \abs{\vp} \abs{\vpp} + mm'}{p\dt p'+ mm'} \right]^{1/2}
   [\cos{\frac{\theta}{2}}\sin{\frac{\theta'}{2}}\xp{-i\phi'}
  -\sin{\frac{\theta}{2}}\cos{\frac{\theta'}{2}}\xp{-i\phi }]
\nonumber \\[12pt]
    C_{--} & = &  C_{++}^*  \nonumber \\[12pt]
    C_{-+} & = & -C_{+-}^*.
\end{eqnarray}
%
The angles $(\theta,\phi)$ and $(\theta',\phi')$ correspond to the axial
and  azimuthal angles for \vp and \vpp respectively.   In the center of
momentum frame where $ \vpp=-\vp$  the expressions for $\cll$  simplify
to,
\begin{eqnarray}
C_{\pm\pm} & = & 0 \nonumber \\[6pt]
C_{\pm\mp} & = & \mp\xp{\pm i \phi}.
\end{eqnarray}
This remains true whether
or not $m=m'$.   If in addition it is also true that $m=m'$ then
(\ref{meqn}) simplifies to
\begin{eqnarray}
u(p',\lambda) \otimes \bar{u}(p,\lambda) & = &
-\lambda\frac{1}{2}\gamma_0(\ps+m)\gamma_5
\rlap /\eta^\lambda \xp{i\lambda\phi} \nonumber
\\[12pt]
u(p',\lambda) \otimes \bar{u}(p,-\lambda) & = &
-\lambda\frac{1}{2}\gamma_0(\ps+m) (1-\lambda\gamma_5 \rlap
/\eta^3)\xp{ i\lambda\phi},
\label{meqn2}
\end{eqnarray}
where
\begin{equation}
\eta^\lambda = \eta_1 - i \lambda \, \eta_2 .
\label{pol12}
\end{equation}
These results also follow by using the relation\cite{haber-94}
\[u(p,\lambda)=-\lambda\gamma_0\, u(-p,-\lambda),\]
in expression (\ref{emb}).

Finally we remark that in the limit $m\rightarrow 0$ equations
(\ref{meqn}) and (\ref{meqn2}) still apply with the proviso that
$\eta^\mu_3\approx p^\mu/m$ + ${\cal O}(m/E)$.  In the massless
limit, where $m',m\rightarrow 0$, expression (\ref{meqn}) simplifies
to
\begin{equation}
u(p',\lambda')\bar{u}(p,\lambda)=
   \left\{ \begin{array}{ll}
         \dfrac{1}{2\sqrt{2p\dt p'} }\, C_{\lambda'\lambda}\,\ps'
          \ps\,(1+\lambda\gamma_5) &
           \mbox{for $\lambda'=-\lambda $}, \\[12pt]
         \dfrac{1}{ 2\sqrt{2p\dt p'} }\, C_{\lambda(-\lambda)}\,
                \ps'\ps\,\gamma_5\rlap /\eta^\lambda &
          \mbox{for $\lambda'=\lambda $}.
            \end{array}\right. \label{meqn3}
\end{equation}
It is easy to verify that (\ref{meqn3}) satisfies all of the identities
valid for the inner product of two massless spinors.  For example,
\[ \bar{u}(p,-)\,u(p',+) = -(\bar{u}(p,+)\, u(p',-))^*, \]
and,
\[ | \bar{u}(p,+)\,u(p',-)|^2 = 2p\cdot p'. \]
Equation (\ref{meqn}), and its simplified forms (\ref{meqn2})
and (\ref{meqn3}), play a central role in our method.
With them we can  evaluate the trace in (3)
and obtain an analytic expression for the amplitude in
terms of dot products.

Evaluating the trace, even at the amplitude level, may
lead to very complicated expressions if the number of external
gauge bosons connected to a single fermion
line is large.  Therefore, we develop an iterative method for obtaining this
trace.  The iterations can be easily performed either symbolically
(using code like REDUCE or FORM) or numerically.
The basic idea is to use the fact that any product of an odd number of
gamma matrices can be written in the form,
\begin{equation}
\dg{B}_1\dg{B}_2\cdots\dg{B}_{2n+1}
= \dg{V}_n + \gamma_5\dg{A}_n.
\end{equation}
Where the $V$'s and $A$'s are obtained iteratively from,
\begin{eqnarray}
   V_n^\mu &=& \frac{1}{4} tr[\gamma^\mu
   (\rlap/V_{n-1}+\gamma_5\dg{A}_{n-1})
     \dg{B}_{2n}\dg{B}_{2n+1}]     \nonumber   \\
   A_n^\mu &=&  \frac{1}{4}
tr[\gamma^\mu(\gamma_5\rlap/{V}_{n-1}+\dg{A}_{n-1})
     \dg{B}_{2n}\dg{B}_{2n+1} ],   \nonumber \\
V_\circ^\mu &=& B_1^\mu  \nonumber   \\
A_\circ^\mu &=& 0. \label{ite}
\end{eqnarray}
This result follows after repeated use of the $\gamma$
identity\footnote{In our conventions $ \gamma_5 = i \gamma^0 \gamma^1
\gamma^2 \gamma^3 $ and $ \epsilon_{ 0 1 2 3 } = + 1 $.},
\[ \gamma_\mu\gamma_\nu\gamma_\sigma =
     g_{\mu\nu}\gamma_\sigma+g_{\nu\sigma} \gamma_\mu
        -g_{\mu\sigma}\gamma_\nu
     + i\gamma_5\eps_{\mu\nu\sigma\rho}\gamma^\rho. \]

For convenience we define the following expressions,
\begin{eqnarray}
F_{\lambda,\lambda'}(V_n,A_n)&=&
   \ub(\dg{V_n} + \gamma_5 \dg{A}_n) \upp \nonumber \\[6pt]
F_{\lambda,\lambda'}(B_0,V_n,A_n)&=&
    \ub\dg{B}_0(\dg{V_n} + \gamma_5 \dg{A_n}) \upp.
\label{mfn}
\end{eqnarray}
The essence of the method is to write expressions of the form
(\ref{one}) in terms of $F_{\lambda,\lambda'}(V_n,A_n)$ and
$F_{\lambda,\lambda'}(B_0,V_n,A_n)$ defined above.
The $V_n$'s and $A_n$'s are obtained by the simple iteration process
(\ref{ite}).  Finally, using (\ref{meqn}) we express (\ref{mfn})
in terms of four vector products.  Since the general expressions for
$F_{\lambda,\lambda'}(V_n,A_n)$ and $F_{\lambda,\lambda'}(B_0,V_n,A_n)$
are common to every Feynman diagram we display their form below.
\begin{eqnarray}
F_{\lambda,\lambda'}(V,A) &=&
 2\,{\cal N} \sum_{\lambda''} C_{\lambda',\lambda''}
M_{\lambda'',\lambda}  \nonumber \\
F_{\lambda,\lambda'}(B_0,V,A) &=&
 2\,{\cal N} \sum_{\lambda''} C_{\lambda',\lambda''}
N_{\lambda'',\lambda}
\end{eqnarray}
where
\begin{eqnarray}
    M_{\lambda'',\lambda} &=&
   \left\{\begin{array}{ll}
        m'p\dt V + mp'\dt V  & {} \\
       \ \ \ \ \ \ \ \ -  \lambda [ mm'\eta^3\dt A
        + \eta^3\dt W(p,p',A,V) ]   &
     \mbox{for $\lambda''=\lambda$} \\[12pt]
     - \left[ mm'e^\lambda\dt A +
    e^\lambda\dt W(p,p',A,V) \right]  & \mbox{for $\lambda''=-\lambda,$}
   \end{array}\right.  \\[18pt]
    N_{\lambda'',\lambda} &=&
   \left\{\begin{array}{ll}
        B_0\cdot W(p,p',V,A) + mm'B_0\cdot V  & {} \\[6pt]
        \ \ \ \ \ +\lambda\, B_0 \cdot W(\eta^3,m' p+ m p',A,V)
        &  \mbox{for $\lambda = \lambda$} \\[12pt]
       B_0 \cdot W(\eta^\lambda,m' p+ m p',A,V)
         & \mbox{for $\lambda'' = -\lambda$,}
   \end{array}\right.
\end{eqnarray}
and,

\begin{equation}
       W^\mu(x,y,z,w) = x^\mu y\dt z + z^\mu x\dt y - y^\mu x\dt z
       + i\epsilon^{\mu\nu\rho\sigma }x_\nu y_\rho w_\sigma\; .
\end{equation}

Of course these expressions take on much simpler forms when
(\ref{meqn2}) and (\ref{meqn3}) are applicable.  We will see an
example of this simplification in the next section.  We now have at
hand all the tools necessary to express any object of the form (3) in
terms of four vector products.  In the next section we illustrate how
these techniques are applied to compute the cross section for various
processes.

\section{Applications}

As an illustration of our method we present the helicity amplitudes for
two processes involving the top quark, namely,
\[e^+e^- \rightarrow t\bar{t} \]
and,
\[ \gamma\gamma \rightarrow t\bar{t}.\]
These processes will be important in the study of the
phenomenology of the top quark at future electron-positron
colliders.

We can write the helicity amplitudes for both
processes in terms of the spinor objects,
\begin{eqnarray}
A_\mu (p,\lambda;p',\lambda')&\equiv&
\bar{u}(p,\lambda)\gamma_\mu\,v(p',\lambda') \nonumber\\
B_\mu(p,\lambda;p',\lambda')&\equiv&
\bar{u}(p,\lambda)\gamma_\mu\gamma_5\,v(p',\lambda')
\end{eqnarray}

We will work in the center of momentum system with the initial
particles moving along the $z$-axis; we also take $ m = m '$.
In this frame we can then use (\ref{meqn2}) to obtain simple
expressions for $A_\mu$ and $B_\mu$,
\begin{equation}
\begin{split}
A_\mu(p,\lambda;p',\lambda') &=
    \begin{cases}
       - 2 m \lambda \, g_{\mu i} \, \hat{p}^{i}&
           \text{if $\lambda'=\lambda$} \\
      -2E\eta_\mu^\lambda(p)& \text{if $\lambda'=-\lambda$},
    \end{cases}
\\[10pt]
%
B_\mu (p,\lambda;p',\lambda') &=
    \begin{cases}
      -2m \,g_{o\mu}& \text{if $\lambda'=\lambda$} ,\\
      -2\lambda\!\abs{\vec{p}}\!\eta_\mu^\lambda
          & \text{if $\lambda'=-\lambda$}.
    \end{cases}
\label{spinfac2}
\end{split}
\end{equation}
Where $\eta_\mu^\lambda(p )$ are the polarization vectors defined in
(\ref{pol12}) corresponding the momentum $p$, mass $m$ and helicity index
$\lambda$.

\begin{figure}
\begin{center}
\mbox{ \includegraphics[scale=.45]{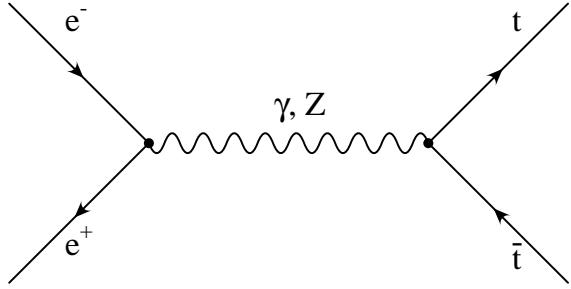} }
\end{center}
\caption{Feynman diagrams for the process $e^+e^-\rightarrow
t\bar{t}$}
\label{fig1}
\end{figure}

\subsection*{\underline{The $e^-e^+ \rightarrow t\bar{t}$ Process:}}
We now have the necessary ingredients to obtain expressions for the
helicity amplitudes.   The Feynman diagrams for the first process
are shown in figure~1.   We take the $\gamma t\bar{t}$, $Z t\bar{t}$,
and  $Ze^-e^+$ couplings to be of the form\footnote{
In the standard model the parameters $q_t,a,b,c$, and $d$ are given by
\begin{alignat}{4}
a & =e\,{ 4 s_w^2 - 1 \over 4 s_w c_w},  & \qquad
c & = e\,{ 1 - 8 s_w^2/3 \over 4 s_w c_w}, & \qquad
b = -d & = { e\over 4 s_w c_w},  & \qquad
q_t & = - \frac{2e}{3}. \notag
\end{alignat}},
\begin{alignat}{2}
\Gamma^\mu_{\gamma t\bar{t}} &= q_t\gamma^\mu,
& \qquad
\Gamma^\mu_{zt\bar{t}} & = \gamma^\mu(c+d\gamma_5)
\\[6pt]
\Gamma^\mu_{\gamma e\bar{e}} & = e\gamma^\mu,
& \qquad
\Gamma^\mu_{ze\bar{e}} & = \gamma^\mu(a+b\gamma_5).
\end{alignat}
%
%
The amplitudes can then be written in terms of (\ref{spinfac2}),
\begin{equation}
\begin{split}
T_\gamma (\lambda_e,\lambda_{\bar{e}};\lambda_t,\lambda_{\bar{t}}) &=
 -\frac{e\,q_t}{s}
A_\mu (\lambda_e,\lambda_{\bar{e}}){}^\dagger\,
A^\mu (\lambda_t,\lambda_{\bar{t}})
\\
T_z(\lambda_e,\lambda_{\bar{e}};\lambda_t,\lambda_{\bar{t}}) &=
 -\frac{1}{s-m_z^2}\biggl\{
 ac\, \bigl[A_\mu (\lambda_e,\lambda_{\bar{e}}){}^\dagger\,
A^\mu (\lambda_t,\lambda_{\bar{t}})\bigr]
\\ &\phantom{{=}\frac{1}{s-m_z^2}\biggl\{}
+ad\, \bigl[A_\mu (\lambda_e,\lambda_{\bar{e}}){}^\dagger\,
B^\mu (\lambda_t,\lambda_{\bar{t}})\bigr]
\\ &\phantom{{=}\frac{1}{s-m_z^2}\biggl\{}
+bc\,\bigl[B_\mu (\lambda_e,\lambda_{\bar{e}}){}^\dagger\,
A^\mu (\lambda_t,\lambda_{\bar{t}})\bigr]
\\ &\phantom{{=}\frac{1}{s-m_z^2}\biggl\{}
+bd\,\bigl[B_\mu (\lambda_e,\lambda_{\bar{e}}){}^\dagger\,
B^\mu (\lambda_t,\lambda_{\bar{t}})\bigr]
\biggr\}.
\end{split}
\label{tamp}
\end{equation}
For simplicity we have suppressed the momentum indices, but the
momentum dependence should be clear from the context.
The electron and positron momenta are designated by $p_e$ and
$p_{\bar{e}}$, and the top, anti-top quark momenta are designated by
$p$ and $\bar{p}$.
In the CM frame the momenta take the form\footnote{These are contravariant four
vectors.},
\begin{align}
p_e &= (E;0,0,E) \notag \\
p_{\bar{e}} &= (E;0,0,-E) \notag \\
p &= (E;\abs{\vec{p}}\cos{\theta},0,\abs{\vec{p}}\sin{\theta})
\notag \\
\bar{p} &= (E;-\vec{p}),  \label{CMvecs}
\end{align}
where $E$ is the energy of the incoming electron in the center of mass frame.
The electron mass is ignored in all of our expressions.  Using these
expressions for the momenta together with
(\ref{spinfac2}) and (\ref{tamp}) it is simple to obtain an expression
for the helicity amplitudes.  We represent the total amplitude by,
$T=T(\lambda_e,\lambda_{\bar{e}};\lambda_t,\lambda_{\bar{t}})$, where
the helicity index for each particle is manifest.  Then,
\begin{equation}
\begin{split}
      T(\lambda_e,-\lambda_{e};\lambda_t,-\lambda_{t}) &=
         -\frac{1}{\beta_z^2}
         \bigl(\,\lambda_e\lambda_t+\cos{\theta}\,\bigr) \\
& \phantom{-\frac{\lambda_e\lambda_t}{\beta_z^2}\bigl(1+}
   \bigl[\,e\,q_t\,\beta_z^2+(a+b\lambda_e)(c+d\,\lambda_t\beta_t)\,\bigr]
\\[6pt]
T(\lambda_e,-\lambda_{e};\lambda_t,\lambda_{t}) &=
  -\frac{m_t\,\lambda_t}{E\,\beta_z^2}\,
      \bigl[ \,e\,q_t\,\beta_z^2 + (a+b\lambda_e)c\, \bigr],
\end{split}
\end{equation}
where $\beta_{z}=\sqrt{1-m_{z}^2/4E^2}$, and $\beta_t=\abs{\vec{p}}/E$.
In obtaining the
above expression the following relations were useful,
\begin{gather}
\vec{\eta}(-\lambda_e,p_e)\cdot\vec{\eta}(\lambda_t,p) =
(\cos{\theta}+\lambda_e\lambda_t) \notag \\
\hat{p}\cdot\vec{\eta}(-\lambda_e,p_e)=\sin{\theta}\notag \\
\hat{p_e}\cdot
\vec{\eta}(-\lambda_e,p_e)\times\vec{\eta}(\lambda_t,p) =
-i(\lambda_t+\lambda_e\cos{\theta}).
\end{gather}

\subsection*{\underline{The $\gamma\gamma \rightarrow t\bar{t}$ Process:}}
As another example we consider the process
$ \gamma \gamma \rightarrow t \bar t $ which will be present in the
back-scattered laser mode of an $ e^+ e^- $ collider. This experiment
will have the possibility of polarizing both photons and of analyzing
the polarization of the final state quarks.

\begin{figure}
\begin{center}
\mbox{\includegraphics[scale=.8]{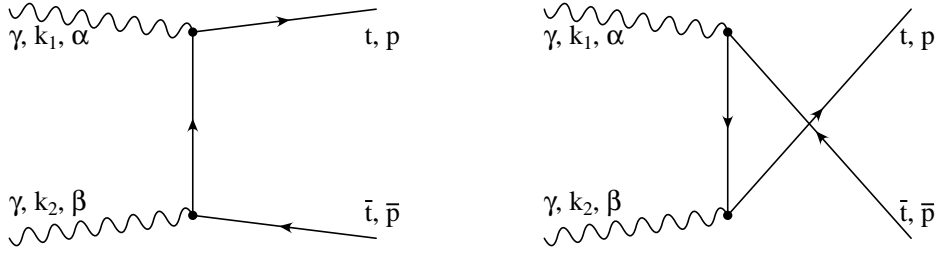} }
\end{center}
\caption{Feynman diagrams for the process
$\gamma\gamma\rightarrow t\bar{t}$}
\label{fig2}
\end{figure}

The relevant diagrams are those obtained via a $t$-quark exchanged in the
$t$-channel (fig.~\ref{fig2}).
\noindent A straightforward application of
(\ref{meqn}) or (\ref{mfn}) in the CM frame (\ref{CMvecs})
leads to the following expression for the amplitude $ T $,
\def\pb{{\bar p}}
\def\lamt{{\lambda_t }}
\def\lamtbar{{ \lambda_{\bar t} }}
\begin{equation}
T(h_1,h_2,\lamt,\lamtbar) =
\frac{q_t^2}{2}\,\left[ { T_{(+)}^{ij }(h_1,h_2,\lamt,\lamtbar) \over p \cdot
k_1 }
+ { T_{(-)}^{ij }(h_1,h_2,\lamt,\lamtbar) \over p \cdot k_2 }
\right] \vartheta_1^i \vartheta_2^j
\end{equation}
where $\lamt,\lamtbar$ denote the top and anti-top helicities and
$ \vartheta_{ 1 , 2 } $ denote the photon polarization vectors,
\begin{equation}
\vec{\vartheta_1} = { 1 \over \sqrt{ 2 } } ( 1 , ih_1 , 0 ) ; \quad
\vec{\vartheta_2} = { 1 \over \sqrt{ 2 } } ( 1 , -ih_2 , 0 ) ,
\end{equation}
denoting by $ h_{ 1 , 2 } = \pm 1$ the photon helicities. We also
defined
\def\ap{ {\abs{ \vec{p} }} }
\begin{equation}
\begin{split}
T_{(\pm)}^{ij }(h_1,h_2,\lamt,\lamt) & =
    2m_t\,\biggl[
\frac{\lamt}{\ap}\bigl(2p^ip^j\pm\vec{k}_1\cdot\vec{p}\,\delta^{ij}\bigl)
     + i\epsilon^{ijl}\,k_1^l \biggl] \\
T_{(\pm)}^{ij }(h_1,h_2,\lamt,-\lamt) & =
    2E\,\biggl[ p^i \left( \eta^\lamt \right)^j +
                p^j \left( \eta^\lamt \right)^i
            \pm\vec{k}_1\cdot\vec{\eta}^\lamt\,\delta^{ij}\biggl]
\end{split}
\end{equation}

In terms of the momenta in (\ref{CMvecs}) we obtain,
\begin{equation}
\begin{split}
T ( h_1,h_1,\lamt , \lamtbar ) & =
q_t^2\,\frac{m_t}{E}\,\frac{2}{1-\beta_t^2\cos^2{\theta}}\,
          \biggl(h_1+\lamt\beta_t\biggr)\,\delta_{\lamt\lamtbar}
\\[18pt]
T ( h_1,-h_1,\lamt , \lamtbar ) & =
 q_t^2\, \frac{2\,\beta_t\,\lamt\sin\theta}{1-\beta_t^2\cos^2{\theta}}
  \begin{cases}
     \dfrac{m_t}{E}\sin{\theta} & \text{if $\lamtbar=\lamt$},\\[10pt]
     (h_1+\lamt\cos{\theta}) & \text{if $\lamtbar=-\lamt$}.
  \end{cases}
\end{split}
\end{equation}
here $ \beta_t $ denotes the velocity of the top-quark, $ \beta_t =
\abs{\vec{p}} / E $, and $ \theta $ is the CM scattering angle.
Note that as $ E \rightarrow \infty $,
$ T ( \lamt , \lamtbar , \pm , \mp ) \rightarrow 0 $ as required by
helicity conservation.  Also in the non-relativistic limit
the amplitude for two photons into two like-spin electrons
vanishes, as it should.

\section{Concluding Remarks}

We have presented a new method for computing helicity amplitudes which
is applicable for processes involving both massless and massive
fermions.  The method allows one to obtain a covariant expression for
the helicity amplitude without having to go through the process of
squaring the amplitude.  The method can be easily adopted to both
numerical and symbolic computations.  We presented two examples which
illustrate the simplicity of the method.  Although for our two
examples we worked in the center of mass frame the method is
applicable in any frame and can be used in processes with many pairs
of external massive fermions.  In particular the methods presented
here were used to compute the cross sections for $gg\rightarrow
t\bar{t}H$, $gg\rightarrow t\bar{t} + b\bar{b}$, $gg\rightarrow
\gamma\gamma t\bar{t}$, and many others (see for example references~
\cite{djv-94,djv-95}).  For very complicated cases the combination of
this method with a program for algebraic manipulation (such as Form,
Reduce, Macyma, Maple, or Mathematica) should greatly simplify
calculations.  Even in those cases where one is only interested in the
total cross section the use of the methods presented here may, in some
cases, prove simpler to use than the traditional approach.

\section{Acknowledgements}
R.V. would like to thank Howard Haber and Michael Peskin for
very helpful discussions.  This research was
supported in part by DOE contracts
DE-FG05-92ER40722 and DE-FG03-92ER40837
and by the Lightner-Sams Foundation of Dallas.

\section{Appendix}

In this appendix we present a proof of the Michel-Bouchiat
identity~(\ref{emb}).  We follow here the notation of
reference~\cite{i&z}.

The state of a particle can be characterized by giving the momentum
$\vec{p}$, mass $m$, total spin $s$, and its component of spin along
a fixed direction.  It is common to take the $z$-axis
or the direction of the momentum to specify this fixed direction.
The latter case is refered to as the helicity basis.  In what follows
we will use the helicity basis.  The advantages of using the helicity
basis are well known\cite{jacob/wick-59} and we will not dwell on
them here.

The helicity operator is defined by,
\[
\Lambda \equiv -\dfrac{2 W\cdot \hat{n}}{m}
\]
where $ W^\mu=\frac{1}{2}\epsilon^{\mu\nu\rho\sigma}P_\nu J_{\rho\sigma}$
is the Pauli-Lubanski four vector; $J_{\mu\nu}$ and $P_{\mu}$
are the angular momentum and  energy-momentum operators respectively.
The commutation relation for the $W$'s is,
\[ [W^\mu,W^\nu]=-i\epsilon^{\mu\nu\rho\sigma}\,p_\rho W_\sigma.\]
This relation follows from the commutation properties of  $J_{\mu\nu}$ and
$P_{\mu}$.
The four vector $n$ is any space-like normalized four vector
with $n^2=-1$.  In the helicity basis $\vec{n}$ is in the direction
of $\vec{p}$.  A conventional form for $n$ is,
\[
n_\mu =
(\frac{|\vec{k}|}{m},\frac{k^0}{m}\frac{k_i}{|\vec{k}|}).
\]

Following Michel\cite{michel} we introduce an orthogonal set
of tensors\footnote{The use of these tensors was first suggested
to one of us by M.E. Peskin.},
$U^\alpha_\mu=({p_\mu}/{m};\eta^1_\mu,\eta^2_\mu,\eta^3_\mu)$
such that,
\begin{eqnarray*}
\betta^i\cdot p &=& 0   \nonumber  \\
\betta^i\cdot \betta^j& = & - \delta^{ij}  \nonumber \\
g^{\mu\nu} U^{\alpha}_{\mu} U^{\beta}_{\nu} &=& g^{\alpha\beta}
\end{eqnarray*}
If we define, $S^\alpha\equiv -{W^\mu U_\mu^\alpha}/{m}$, then
it follows that, $W_\mu=S^i\eta^i_\mu$. For a particle of spin $s$,
with $\betta\in p\cdot \betta=0$,
the operator, $W\cdot\betta$, has $(2s+1)$ eigenvalues:
$\lambda = \{-s,-s+1,\ldots, s-1,s\}$.   These eigenvalues
are used to label states of given momenta and spin.

 From the commutation relation for the $W$'s and the properties
of the $\eta$'s it is straightforward to show that the
commutation relation for the $S$'s obey an $SU(2)$ algebra,
\[ [S_i,S_j]= i\epsilon_{ijk}S_k, \]
Furthermore, if we pick $\eta^3$ to have the form,
\[
\eta^3_\mu =
\left(
\frac{\vec{p}}{m},\frac{k^0}{m}\frac{k_i}{|\vec{k}|}
\right),
\]
then up to a factor of $\frac{1}{2}$, the operator $S_3$,
is just the helicity operator.
In the helicity basis states are labeled by the eigenvalues of $S_3$.
The helicity projection operator is given by,
\begin{eqnarray*}
\Lambda_{\pm} =
& \frac{1}{2}(1 \mp {W\cdot \betta_3}/{m})
\nonumber \\[6pt]
= & \frac{1}{2} \left(1\pm 2S_3\right),
\end{eqnarray*}
and as usual the lowering and raising operators of helicity are given by,
\[ S_\pm = S_1 \pm i S_2. \]

We label the states for particles of spin $s=\frac{1}{2}$
by $u(p,\lambda)$ where,
\begin{eqnarray}
2S_3\, u(p,\lambda) = & \lambda\, u(p,\lambda)
\nonumber \\
\ps\, u(p,\lambda) = & m\, u(p,\lambda).
\end{eqnarray}
Then,
\begin{equation}
u(p,\pm)\,\bar{u}(p,\pm) =
\dfrac{(1 \pm 2 S_3)}{2}(\ps+m),
\end{equation}
where we are using a normalization such that,
$\bar{u}(p,\lambda) u(p,\lambda)=2m$.

In order to generalize the above expression to the case where the spinors
have different helicity we simply use the raising and lowering operator
for helicity,
\begin{eqnarray*}
u(p,\mp)\,\bar{u}(p,\pm)
& = & S_\mp\, u(p,\pm)\,\bar{u}(p,\pm) \nonumber \\
& = & S_\mp\, \frac{( 1 \pm 2 S_3 )}{2}( \ps + m ) \nonumber \\
& = & ( S_1 \mp i S_2 )( \ps+m ),
\end{eqnarray*}
where we have used,
$S_\mp\, S_3 = \pm\frac{1}{2} S_\mp$.

To obtain the Michel-Bouchiat form (\ref{emb}) we recall that,
$S_i=-{W \cdot\eta_i}/m$ and use the definition of $W_\mu$
to obtain,
\[ S_i=\frac{\gamma_5\rlap/ \eta_i\ps}{2m},\]
and,
\begin{eqnarray*}
u(p,\pm)\,\bar{u}(p,\pm)
& = &  \frac{1 \pm  \gamma_5\rlap/ \eta_3}{2} (\ps+m), \\[6pt]
 u(p,\mp)\,\bar{u}(p,\pm)
& = &  \gamma_5\frac{\rlap/ \eta_1 \mp i\rlap/ \eta_2}{2}
(\ps+m).
\end{eqnarray*}
Finally using the Pauli matrices, $\sigma^i$, these expressions
can be written in the shorthand notation of (\ref{emb}).

\end{document}